\documentclass[12pt]{iopart}

\usepackage{graphicx}
\usepackage[colorlinks=true,linkcolor=blue,citecolor=blue,filecolor=cyan,urlcolor=blue,breaklinks=true]{hyperref}
\usepackage[utf8]{inputenc}
\usepackage{cite}

\begin{document}

\title[]{Thermodynamic bound on current fluctuations in coherent conductors}

\author{Kay Brandner\textsuperscript{1,2}, Keiji Saito\textsuperscript{3}}

\address{\textsuperscript{1} School of Physics and Astronomy, University of Nottingham,
Nottingham NG7 2RD, United Kingdom\\
\textsuperscript{2} Centre for the Mathematics and Theoretical Physics of Quantum Non-Equilibrium Systems, University of Nottingham, Nottingham NG7 2RD, United Kingdom\\
\textsuperscript{3} Department of Physics, Kyoto University, Kyoto 606-8502, Japan
}

\ead{kay.brandner@nottingham.ac.uk}

\vspace{10pt}
\begin{indented}
\item[] \today
\end{indented}

\begin{abstract}
We derive a universal bound on the large-deviation functions of particle currents in coherent conductors.
This bound depends only on the mean value of the relevant current and the total rate of entropy production required to maintain a non-equilibrium steady state, thus showing that both typical and rare current fluctuations are ultimately constrained by dissipation.
Our analysis relies on the scattering approach to quantum transport and applies to any multi-terminal setup with arbitrary chemical potential and temperature gradients, provided the transmission coefficients between reservoirs are symmetric.
This condition is satisfied for any two-terminal system and, more generally, when the dynamics of particles within the conductor are symmetric under time-reversal.
For typical current fluctuations, we recover a recently derived thermodynamic uncertainty relation for coherent transport.
To illustrate our theory, we analyze a specific model comprising two reservoirs connected by a chain of quantum dots, which shows that our bound can be saturated asymptotically.
\end{abstract}

%\maketitle

\newcommand{\kb}{k_\mathrm{B}}

\newcommand{\eqref}[1]{(\ref{#1})}

\newcommand{\bA}{\mathbf{A}}
\newcommand{\bB}{\mathbf{B}}
\newcommand{\bC}{\mathbf{C}}
\newcommand{\bD}{\mathbf{D}}
\newcommand{\bF}{\mathbf{F}}
\newcommand{\bS}{\mathbf{S}}
\newcommand{\bT}{\mathbf{T}}
\newcommand{\bV}{\mathbf{V}}
\newcommand{\bZ}{\mathbf{Z}}

\newcommand{\cF}{\mathcal{F}}
\newcommand{\cT}{\mathcal{T}}

\newcommand{\bx}{\biggl[}
\newcommand{\by}{\biggr]}

\renewcommand{\a}{\alpha}
\renewcommand{\b}{\beta}
\newcommand{\g}{\gamma}
\renewcommand{\d}{\delta}

\newcommand{\qu}{\mathrm{qu}}
\newcommand{\cl}{\mathrm{cl}}
\newcommand{\arcoth}{\mathrm{arcoth}}

\renewcommand{\tr}{\mathrm{Tr}}
\newcommand{\Eint}{\int_0^\infty \! dE \;}

\newcommand{\ket}[1]{|#1 \rangle}
\newcommand{\bra}[1]{\langle #1 |}

\section{Introduction}

Mesoscopic conductors  provide a powerful platform to realize quantum devices like single-electron transistors \cite{fujiwara2008,bitton2017,asgari2021}, electronic interferometers \cite{ji2003,jo2021,chakraborti2025}, or thermoelectric heat engines and refrigerators \cite{prance2009,brantut2013,josefsson2018}.
At the same time, they form a versatile testbed to explore the interplay between statistical physics and quantum mechanics under non-equilibrium conditions, and under the influence of thermal and quantum fluctuations, which play a dominant role on small length end energy scales \cite{seifert2012,benenti2017,bauerle2018}. 
These fluctuations render thermodynamic fluxes like the amount of particles or heat absorbed by an external reservoir into stochastic variables, which are no longer fully characterized by their mean values. 
Instead, they need to be described through distribution functions, which are formally defined as 
\begin{equation}
	P_t(A) = \bigl\langle\delta[A-\hat{A}_t]\bigr\rangle. 
\end{equation}
Here, $\hat{A}_t$ is the Heisenberg-picture operator that represents the observable $A$ and angular brackets denote the average over all admissible quantum states of the system.  
Understanding the universal features of these distributions has emerged as a central problem in mesoscopic physics over the past decades.
This research has led to a range of fundamental insights, which can be broadly divided into two groups: 
equalities, like fluctuation theorems, which describe global properties of flux distributions and originate from symmetries of the underlying microscopic dynamics \cite{andrieux2006,andrieux2007,saito2008,nakamura2010,sanchez2010,hussein2014,gaspard2015}; and inequalities that bound the first and second cumulants of these distributions in terms of entropy production, dynamical activity or other physical quantities \cite{potanina2021,eriksson2021,tesser2023,tesser2024,acciai2024,palmqvist2024,palmqvist2025,blasi2025,brandner2025c}. 
While the first class of relations provides stronger constraints on the structural details of flux distributions, the bounds of the second class tend to have greater predictive power. 
Thermodynamic uncertainty relations, for example, connect the total entropy production incurred by a transport process with the mean and variance of every individual flux entering the conductor \cite{seifert2019,horowitz2020}. 
As a result, it becomes possible to bound the overall dissipation rate, which may arise from a multitude of different fluxes, some of which may not be measurable, in terms of observables that can be easily accessed in experiments, a strategy known as thermodynamic inference \cite{seifert2019,horowitz2020}. 
Furthermore, thermodynamic uncertainty relations can be used to derive trade-off relations between the figures of merit of small-scale devices, such as the power, efficiency and constancy of mesoscopic heat engines and refrigerators \cite{pietzonka2018,koyuk2019}.
Such relations are of significant conceptual and practical value as they reveal fundamental limits on device performance and provide universal benchmarks for both theoretical models and experimental realizations. 

Neither of these applications would be possible based solely on equality relations like fluctuation theorems, which do in general not establish strong links between flux statistics and thermodynamic quantities like entropy production.
At the same time, however, most of the currently available bounds on fluctuations do not contain any information about the corresponding distributions beyond their second cumulant.
Large-deviation theory offers a powerful tool to bridge this gap. 
In essence, this approach relies on the observation that, upon applying a suitable rescaling, the time dependence of steady-state flux distributions typically becomes generic after a short transient period. 
More precisely, the distribution $p_t(j_A) = t P_t(j_A t)$ of the current $j_A=A/t$, which, in contrast to the flux $A$ is not extensive in time, takes the general form  
\begin{equation}
	p_t(j_A) = \exp[-I(j_A) t + o(t)],
\end{equation}
where $o(t)$ stands for corrections that grow strictly slower than $t$ in the limit $t\rightarrow\infty$. 
Thus, at sufficiently long times, the shape of the distribution $p_t(j_A)$ is dominated by the rate function $I(j_A)$, which is usually convex and attains its minimum at the mean value of the current, 
\begin{equation}
	J_A = \lim_{t\rightarrow\infty} \frac{\langle \hat{A}_t\rangle}{t}.
\end{equation}
The variance, or noise, of the current is determined by the curvature of $I(j_A)$ at $j_A=J_A$ through the relation 
\begin{equation}\label{eq:int:DefCN}
	S_A = \lim_{t\rightarrow\infty} 
			\frac{\langle\hat{A}_t^2\rangle-\langle\hat{A}_t^{\vphantom{2}}\rangle^2}{t}
		= \frac{1}{I''(J_A)},	
\end{equation}
and similar formulas can be derived for higher-order cumulants. 
However, the actual merit of the rate function $I(j_A)$ lies in the fact that it characterizes the full current distribution beyond the level of individual cumulants. 
In particular, it accurately captures also large deviations from the mean value, which occur with exponentially small probability and cannot be easily described by other means \cite{touchette2009a,jack2020,burenev2025}. 
As a result, it becomes possible to systematically investigate whether not only typical but also rare current fluctuations are constrained by universal dynamic or thermodynamic principles.

For the classical regime, where the microscopic dynamics of mesoscopic conductors can be modeled with time-homogeneous Markov jump processes, the answer to this question is indeed affirmative. 
In such a setting, the rate function of any individual current admits the remarkably simple bound 
\begin{equation}\label{eq:int:LDFBndCl}
	I(j_A) \leq \hat{I}_\cl (j_A) 
		= \frac{(j_A- J_A)^2}{4\kb J_A^2}\sigma,
\end{equation}
where $\sigma$ is the total rate of entropy production that is incurred by maintaining the system in a non-equilibrium steady state, and $\kb$ denotes Boltzmann's constant \cite{pietzonka2016,gingrich2016}. 
This result implies the standard thermodynamic uncertainty relation 
\begin{equation}\label{eq:int:TURCl}
	\mathcal{Q}_\cl = \frac{S_A\sigma}{2\kb J_A^2} \geq 1,
\end{equation}
which shows that the relative uncertainty $S_A/J_A^2$ of any current is ultimately bounded by the overall dissipation rate of the transport process \cite{seifert2019,horowitz2020}. 
Moreover, Eq.~\eqref{eq:int:LDFBndCl} entails that this principle also applies to large fluctuations, the likelihood of which cannot be arbitrarily suppressed without generating more entropy. 
These conclusions hold arbitrarily far from equilibrium and regardless of the internal structure the conductor. 
They can, however, not be easily extended to the quantum regime, where universal and operationally accessible bounds on current fluctuations are still relatively scarce, despite significant efforts to close this gap, see for instance Refs.~\cite{macieszczak2018, carollo2019, guarnieri2019, hasegawa2020, hasegawa2021, rignon-bret2021, menczel2021, vanvu2022, prech2023, vu2024, moreira2024, kwon2024}. 

To make progress in this direction, we here focus on coherent transport processes, which typically emerge when the mean free path of particles becomes comparable to the dimensions of the conductor. 
Such systems can be realized, for example, with semiconductor nano-structures \cite{vanwees1988,wharam1988,schwab2000,matthews2014}, atomic junctions  \cite{krans1995,vandenbrom1999,cui2017,lumbroso2018} or ultracold atomic gases in optical potentials \cite{brantut2012,krinner2015,lebrat2018}.
Besides their practical relevance, coherent conductors offer the advantage of admitting a simple and physically transparent theoretical description in terms of single-particle scattering amplitudes \cite{sivan1986,buttiker1992}. 
Within this framework, it was shown early on that a combination of energy filtering and Pauli blocking makes it possible to generate arbitrarily strong violations of the classical uncertainty relation \eqref{eq:int:TURCl}, and thus the more general bound \eqref{eq:int:LDFBndCl} \cite{brandner2018,saryal2019,ehrlich2021,gerry2022,timpanaro2025}. 
At least for particle currents, which we indicate by omitting the label $A$, this effect can be accounted for solely by adjusting the relative weight of current noise and dissipation. 
Specifically, the generalized thermodynamic uncertainty relation 
\begin{equation}\label{eq:int:TURQu}
	\mathcal{Q}_\qu = \frac{S}{J}\sinh\bx\frac{\sigma}{2\kb J}\by\geq 1
\end{equation}
holds for any coherent multi-terminal conductor and arbitrarily strong chemical and thermal biases, as long as the transmission coefficients that describe the exchange of particles between different reservoirs are symmetric \cite{brandner2025c}. 
The main objective of the present work is to show that this bound can be consistently extended to large fluctuations. 
As our central result, we find that the corresponding rate function $I(j)$ is subject to the universal thermodynamic constraint
\begin{equation}\label{eq:int:LDFBndQu}
	I(j) \leq \hat{I}_\qu(j) = (JR+j)\ln\bx\frac{JR+ j}{JR+J}\by
		+ (JR-j)\ln\bx\frac{JR - j}{JR-J}\by,
\end{equation} 
where the dissipation factor $R$ is defined as 
\begin{equation}\label{eq:int:DissFact}
	R= \coth\bx\frac{\sigma}{4\kb J}\by. 
\end{equation}
This bound, the main physical implications of which are analogous to those of its classical counterpart \eqref{eq:int:LDFBndCl}, applies to the range $-JR < j < JR$ of the fluctuating current $j$ and under the same conditions as the quantum uncertainty relation \eqref{eq:int:TURQu}. 
In fact, Eq.~\eqref{eq:int:TURQu} follow from Eq.~\eqref{eq:int:LDFBndQu}, as can be easily verified by using the general relation \eqref{eq:int:DefCN} and noting that the inequality \eqref{eq:int:LDFBndQu} can only hold if $I''(J)\geq \hat{I}''_\mathrm{qu}(J)$, since $I(j)$ and $\hat{I}_\mathrm{qu}(j)$ both have their global minimum at $j=J$. 

Three remarks are in order before we move on to more technical aspects.
First, the quantum bounds \eqref{eq:int:TURQu} and \eqref{eq:int:LDFBndQu} are weaker than their classical counterparts \eqref{eq:int:TURCl} and \eqref{eq:int:LDFBndCl}, since $\mathcal{Q}_\mathrm{cl}\leq \mathcal{Q}_\mathrm{qu}$ and $\hat{I}_\mathrm{cl}(j)\leq \hat{I}_\mathrm{qu}(j)$ whenever $-JR < j < JR$, as we will show in the next section.  
Second, the scattering formalism that underpins our analysis relies on the assumption that particles do not interact with each other or with external degrees of freedom like impurities or lattice vibrations \cite{sivan1986,buttiker1992}.
Nonetheless, we expect the bound \eqref{eq:int:LDFBndQu} to be robust against moderate perturbations of this kind, as long as they can be described through virtual reservoirs, whose chemical potentials and temperatures are adjusted such that they do not exchange any particles or heat with the conductor on average. 
Such probe terminals provide a convenient tool to account for dephasing and inelastic collisions on a mean-field level, and their inclusion does not compromise the validity of our approach, which applies to arbitrary multi-terminal settings \cite{buttiker1988}. 
Finally, we will demonstrate in the following that the bounds \eqref{eq:int:TURQu} and \eqref{eq:int:LDFBndQu} can be simultaneously saturated for two-terminal conductors with a narrow boxcar-shaped transmission function. 
Quite naturally, this setup also gives rise to maximal violations of the classical uncertainty relation \eqref{eq:int:TURCl}, as has been pointed out earlier \cite{ehrlich2021,timpanaro2025}. 

We proceed as follows. 
In the next section, we outline the mathematical framework that underpins our analysis and provide detailed proofs of the results described above. 
We then illustrate our theory by working out a simple yet instructive example in Sec.~\ref{sec:exp}, which we also use to further discuss the physical origins of our bounds. 
The final Sec.~\ref{sec:per} provides a brief summary of our main insights and perspectives for future research.

\section{Approach}\label{sec:apr}

\subsection{Setup}
We begin with a brief recap of the scattering approach to coherent transport. 
As a generic model for a mesoscopic conductor, we consider a sample connected to $L$ thermal reservoirs via ideal semi-infinite leads, each supporting a maximum of $M_E$ transversal modes at a given energy $E$. 
These leads are introduced solely as a theoretical tool to enable the definition wave functions with scattering boundary conditions. 
The transition of a particle with energy $E$ through the conductor is described by the scattering matrix $\bS_E$, which encodes the physical properties of the sample. 
This matrix can be decomposed into $L^2$ quadratic blocks $(\bS_E)_{\a\b} = \bS^{\a\b}_E$ of size $M_E$, which contain the amplitudes for transitions from the transversal modes of the lead $\b$ to those of the lead $\a$. 
Transport is generated by the reservoirs injecting beams of Fermionic particles with a thermal energy distribution. 
After passing through the conductor, these particles are absorbed again into one of the reservoirs, where we assume that no reflections take place at the boundaries between leads and reservoirs. 
Provided that interactions between particles and inelastic scattering events involving additional degrees of freedom can be neglected inside the conductor, the mean currents of particles and heat that flow from the reservoir $\a$ into the sample then admit the simple expressions \cite{sivan1986,buttiker1992}
\begin{eqnarray}
	\label{eq:apr:DefCurrJ}
	J^\a & = \frac{1}{h}\Eint\sum_{\b}\cT^{\a\b}_E (f^\a_E - f^\b_E),\\
	\label{eq:apr:DefCurrQ}
	J^\a_Q & = \frac{1}{h}\Eint\sum_{\b}\cT^{\a\b}_E (f^\a_E - f^\b_E)(E-\mu_\a).
\end{eqnarray}
Here, $h$ denotes Planck's constant and the transmission coefficients and matrices are defined as 
\begin{equation}\label{eq:apr:DefTrans}
	\cT^{\a\b}_E = \Tr [\bT^{\a\b}_E] \quad\mathrm{and}\quad 
		\bT^{\a\b}_E = \bS^{\a\b}_E (\bS^{\a\b}_E)^\dagger,
\end{equation}
where the trace indicates summation over all transversal modes. 
Thermodynamics enters these formulas via the Fermi functions 
\begin{equation}
	f^\a_E = \frac{1}{1+\exp[(E-\mu_\a)/\kb T_\a]},
\end{equation}
where $\mu_\a$ and $T_\a$ are the chemical potential and temperature of the reservoir $\a$. 

Since the propagation of particles through the conductor is a coherent process described by Schr\"odinger's equation, the only source of dissipation is the thermalization of particles inside the reservoirs. 
The steady-state rate of entropy production is therefore given by
\begin{equation}\label{eq:apr:TotEntProd}
	\sigma = -\sum_\a J^\a_Q/T_\a = \frac{\kb}{h}\Eint\sum_{\a\b} \cT^{\a\b}_E
		(f^\a_E - f^\b_E)\ln[f^\a_E/(1-f^\a_E)],
\end{equation}
where we have inserted the formula \eqref{eq:apr:DefCurrQ} for the average heat currents $J^\a_Q$. 
Using this expression and the sum rules $\sum_\a \cT^{\a\b}_E = \sum_\a \cT^{\b\a}_E$, which reflect the physical principle of current conservation and formally follow from the fact that the scattering matrix $\bS$ is unitary, it can be easily shown that $\sigma\geq 0$ \cite{nenciu2007}. 
Hence, the scattering approach is inherently consistent with the second law of thermodynamics. 
For later purposes, we note that, if the transmission coefficients satisfy the additional symmetry relation $\cT^{\a\b}_E = \cT^{\b\a}_E$, which is the case for any two-terminal system, or if the dynamics inside the conductor are invariant under time reversal \cite{sivan1986,buttiker1992}, any given reservoir $\a$ can be assigned a non-negative partial rate of entropy production
\begin{equation}\label{eq:apr:PartEntProd}
	\sigma_\a = \frac{\kb}{h}\Eint \sum_{\b\neq\a} \cT^{\a\b}_E
		(g^{\a\b}_E - g^{\b\a}_E)\ln[g^{\a\b}_E/g^{\b\a}_E] \leq \sigma,
\end{equation}
where $g^{\a\b}_E = f^{\a}_E(1-f^\b_E)$. 

\subsection{Full Counting Statistics}

As outlined above, the scattering approach yields explicit formulas for the total dissipation rate and the mean values of all particle and heat currents entering the conductor. 
Similar expressions are also available for the variances of these currents, from which the generalized thermodynamic uncertainty relation \eqref{eq:int:TURQu} can be derived directly using a convexity argument \cite{brandner2025c}. 
However, the rate functions of the currents, which characterize the full structure of their distributions at sufficiently long times, are in general not accessible explicitly. 
To circumvent this problem, we return to the flux distribution $P^\a_t(N)$, that is, the probability that $N$ particles have been extracted from the reservoir $\a$ at the time $t$.  
From this distribution, we can construct the moment generating function 
\begin{equation}
	K^\a_t(s) = \sum_{N=-\infty}^\infty P^\a_t (N) e^{s N},
\end{equation}
which, similar to the current distribution $p^\a_t(j)$, takes the generic form 
\begin{equation}
	K_t^\a(s) = \exp[\chi_\a(s)t + o(t)]. 
\end{equation}
Hence, the shape of $K^\a_t(s)$ is dominated by the scaled cumulant generating function $\chi(s)$ at long times, where $s$ is a real counting field. 
This function is usually concave and vanishes at $s=0$. 
Moreover, it is connected to the rate function $I_\a(j)$ of the particle current injected by the reservoir $\a$ through the Legendre-Fenchel transformation \cite{touchette2009a,jack2020,burenev2025}
\begin{equation}\label{eq:apr:LFTransform}
	I_\a(j) = \sup_{s} [js - \chi_\a(s)]. 
\end{equation}
Hence, any lower bound on $\chi_\a(s)$ implies an upper bound on $I_\a(j)$. 
The main benefit of this insight, which underpins the following analysis, is that, in contrast to the rate function, the scaled cumulant generating function does in fact admit and explicit expression, which is given by the Levitov-Lesovik formula \cite{levitov1993,lesovik2011}
\begin{equation}\label{eq:apr:LLForm}
	\chi_\a(s) = \frac{1}{h}\Eint \ln\Bigl[\det\Bigl[1-\bF_E
		+ \bF_E\bS^\dagger_E e^{-s\bV_\a}\bS_E e^{s\bV_\a} \Bigr]\Bigr].
\end{equation}
Here, $\bF_E$ is a diagonal matrix containing the Fermi functions of the reservoirs in the corresponding mode sectors and $\bV_\a$ is the projector onto the mode space of the lead $\a$. 
That is, if these matrices are partitioned conformally with the scattering matrix $\bS_E$, their block entries are given by $(\bF_E)_{\a\b} = \delta_{\a\b} f^\a_E\mathbf{1}_E$ and $(\bV_\a)_{\b\g} = \delta_{\a\b}\delta_{\a\g}\mathbf{1}_E$, where $\mathbf{1}_E$ is the identity matrix with size $M_E$. 
We note that, for the sake of convenience, we notationally suppress the implicit dependence of $\bV_\a$ on energy through the maximum number of accessible transversal modes $M_E$, which determines the dimensions of its non-zero block entry. 

In the next section, we will show that the scaled cumulant generating function of any particle current can be bounded solely in terms of the corresponding mean current and the total dissipation rate. 
For this purpose, it is useful to express the formula~\eqref{eq:apr:LLForm} in terms of the transmission matrices defined in Eq.~\eqref{eq:apr:DefTrans}, which can be achieved as follows. 
We first note that $e^{\pm s\bV_\a} = \mathbf{1}_E + (e^{\pm s} -1)\bV_\a$, since $\bV_\a$ is a projector.  
Using this identity, the determinant in Eq.~\eqref{eq:apr:LLForm} can be rewritten as 
\begin{equation}\label{eq:apr:LLFormAux}
	\det\Bigl[1-\bF_E+ \bF_E\bS^\dagger_E e^{-s\bV_\a}\bS_E e^{s\bV_\a}\Bigr]
	= \det\Bigl[\bA_E^\a + \bB_E^\a \bC_E^\a\Bigr]
\end{equation}
with 
\begin{eqnarray}
	\bA_E^\a & = 1 + \bF_E \bV_\a(e^s-1), \\
	\bB_E^\a & = \bF_E \bS^\dagger_E \bV_\a (e^{-s}-1), \\
	\bC_E^\a & = \bV_\a \bS_E + \bV_\a \bS_E \bV_\a (e^s-1).
\end{eqnarray}
If these matrices are partitioned conformally with the scattering matrix, $\bA_E^\a$ is block diagonal, while $\bB_E^\a$ and $\bC_E^\a$ contain non-zero blocks only in the column and row $\a$, respectively. 
Hence, upon applying the matrix determinant lemma, which states that 
\begin{equation}
	\det [\bA +\bB\bC] = \det[\bA]\det[1+\bC\bA^{-1}\bB]
\end{equation}
for any invertible square matrix $\bA$ and any $\bB$ and $\bC$ of compatible size, the expression \eqref{eq:apr:LLFormAux} simplifies considerably, since the matrix $\bC_E^\a(\bA_E^\a)^{-1}\bB_E^\a$ contains only one non-zero block on its diagonal. 
With the help of the relation 
\begin{equation}
	\sum_\b \bS^{\a\b}_E (\bS^{\g\b}_E)^\dagger = \delta_{\a\g}\mathbf{1}_E,
\end{equation}
which follows from the fact that $\bS_E$ is unitary, the formula \eqref{eq:apr:LLForm} can thus be brought into the form
\begin{equation}\label{eq:apr:LLFormT}
	\chi_\a(s) = \frac{1}{h}\Eint \ln\Bigl[\det\Bigl[1 + \sum_{\b\neq\a} \bT^{\a\b}_E\Bigl(
		g^{\a\b}_E(e^s-1) + g^{\b\a}_E(e^{-s}-1)\Bigr)\Bigr]\Bigr],
\end{equation}
where we have again used the definition $g^{\a\b}_E = f^{\a}_E(1-f^\b_E)$.  

\subsection{Thermodynic Constraint}

The thermodynamic constraint \eqref{eq:int:LDFBndQu} follows from a lower bound on the scaled cumulant generating function \eqref{eq:apr:LLFormT} in terms of the mean particle current \eqref{eq:apr:DefCurrJ} and the partial rate of entropy production \eqref{eq:apr:PartEntProd}, both of which depend only on the transmission coefficients $\cT^{\a\b}_E=\Tr[\bT^{\a\b}_E]$, rather than the full transmission matrices $\bT^{\a\b}_E$. 
To derive this bound, we therefore need to trace out the transversal modes of the leads, which can be achieved by rewriting Eq.~\eqref{eq:apr:LLFormT} as 
\begin{equation}
	\chi_\a(s) 	= \frac{1}{h}\Eint \Tr\Bigl[\phi\Bigl(
		\bZ^\a_E\bD^\a_E (\bZ^\a_E)^\dagger\Bigr)\Bigr]. 
\end{equation}
Here, we have used that the identity $\ln[\det[\bA]] = \Tr[\ln[\bA]]$ holds for any non-singular square matrix $\bA$. 
Furthermore, we have defined the function $\phi(x) = \ln[1+x]$ and the matrices $\bZ^\a_E = \bS_E\bV_\a$ and $\bD^\a_E$, which consist of the blocks 
\begin{eqnarray}
	(\bZ^\a_E)_{\b\g} &= \delta_{\a\g}\bS^{\b\a}_E,\\
	(\bD^\a_E)_{\b\g} &= \delta_{\b\g}(1-\delta_{\a\b})\Bigl(
		g^{\a\b}_E(e^s-1) + g^{\b\a}_E(e^{-s}-1)\Bigr)\mathbf{1}_E. 
\end{eqnarray}
We now observe that, first, $\phi(x)$ is operator concave and vanishes at $x=0$, second, the matrix $\bD^\a_E$ is Hermitian, and third, the spectral norm of $\bZ^\a_E(\bZ^\a_E)^\dagger$ is given by $\Vert \bZ^\a_E(\bZ^\a_E)^\dagger\Vert = \Vert \bV_\a\Vert = 1$.
Under these conditions, we can apply Jensen's trace inequality, which yields the relation \cite{hansen1982}, 
\begin{equation}
	\Tr\Bigl[\phi\Bigl(\bZ^\a_E\bD^\a_E (\bZ^\a_E)^\dagger\Bigr)\Bigr]
	\geq \Tr\Bigl[\bZ^\a_E\phi(\bD^\a_E )(\bZ^\a_E)^\dagger\Bigr].
\end{equation}
Upon reinserting the definitions of $\phi(x)$, $\bZ^\a_E$ and $\bD^\a_E$, we thus obtain the bound 
\begin{equation}\label{eq:apr:SCFBnd1}
	\chi_\a(s) \geq \frac{1}{h}\Eint \sum_{\b\neq\a} \cT^{\a\b}_E
		\ln\Bigl[g^{\a\b}_E(e^s-1) + g^{\b\a}_E(e^{-s}-1)\Bigr]
\end{equation}
on the scaled cumulant generating function, which indeed depends only on the transmission coefficients $\cT^{\a\b}_E$.

From here, we can proceed with a convexity argument similar to the one used in Ref.~\cite{brandner2025c} to derive the generalized thermodynamic uncertainty relation \eqref{eq:int:TURQu}. 
To prepare this step, we note that 
\begin{equation}\label{eq:apr:logBnd}
	\ln[1+x-y] \geq y\ln[x/y]
\end{equation}
for any $x>0$ and any $1>y>0$, as can be easily verified be recalling that the natural logarithm can be expressed as 
\begin{equation}
	\ln[x] = \int_0^\infty \! d\mu\; \frac{x-1}{(1+\mu)(x+\mu)}
\end{equation}
for any $x>0$. 
Using the relation \eqref{eq:apr:logBnd} and the definition $h^{\a\b}_E = [g^{\a\b}_E]^\frac{1}{2}$, we have 
\begin{eqnarray}
	&\ln\Bigl[g^{\a\b}_E(e^s-1) + g^{\b\a}_E(e^{-s}-1)\Bigr]\\
	&\qquad = \ln\Bigl[1+\Bigl(h^{\a\b}_E e^{s/2} + h^{\b\a}_E e^{-s/2}\Bigr)^2
		-\Bigl(h^{\a\b}_E + h^{\b\a}_E\Bigr)^2\Bigr]\nonumber\\
	&\qquad \geq 2 \Bigl(h^{\a\b}_E + h^{\b\a}_E\Bigr)^2
		\ln\bx\frac{h^{\a\b}_E e^{s/2} + h^{\b\a}_E e^{-s/2}}{
		 h^{\a\b}_E + h^{\b\a}_E}\by,\nonumber
\end{eqnarray}
since $0<h^{\a\b}_E + h^{\b\a}_E <1$ any finite $E$. 
This inequality can further be rewritten as
\begin{eqnarray}
	& \ln\Bigl[g^{\a\b}_E(e^s-1) + g^{\b\a}_E(e^{-s}-1)\Bigr]\\
	&\qquad	\geq 2(g^{\a\b}_E - g^{\b\a}_E)\coth\bx\frac{1}{4X^{\a\b}_E}\by
				\ln\bx\frac{\cosh[s/2 + 1/4 X^{\a\b}_E]}{\cosh[1/4 X^{\a\b}_E]}\by
				\nonumber 
\end{eqnarray}
with $X^{\a\b}_E = 1/\ln[g^{\a\b}_E/g^{\b\a}_E]$. 
Inserting this result into Eq.~\eqref{eq:apr:SCFBnd1} yields the new bound 
\begin{equation}\label{eq:apr:SCFBnd2}
	\chi_\a(s) \geq \frac{2\sigma_\a}{\kb} \Eint \sum_{\b\neq\a}
		\cT^{\a\b}_E \Sigma_E^{\a\b}\Phi(X^{\a\b}_E)
\end{equation}
on the scaled cumulant generating function, where  
\begin{eqnarray}
	\Sigma^{\a\b}_E & = \frac{\kb(g^{\a\b}_E - g^{\b\a}_E)}{
		h\sigma_\a}\ln[g^{\a\b}_E/g^{\b\a}_E],\\ 
	\Phi(x) & = x\coth\bx\frac{1}{4x}\by
		\ln\bx\frac{\cosh[s/2+1/4x]}{\cosh[1/4x]}\by
\end{eqnarray}
and $\sigma_\a$ denotes the partial rate of entropy production defined in Eq.~\eqref{eq:apr:PartEntProd}.
It remains to observe that the function $\Phi(x)$ is convex over the entire real axis for any real value of $s$, as we show in \ref{app:SI}. 
Therefore, the relation 
\begin{equation}
	\Phi(X^{\a\b}_E) \geq \Phi(X_\a) + \Phi'(X_\a)(X^{\a\b}_E - X_\a)
\end{equation}
holds for any real $X_\a$, where $\Phi'(x)$ denotes the derivative of $\Phi(x)$. 
Since $\Sigma^{\a\b}_E\geq 0$, this result implies a whole family of lower bounds on the scaled cumulant generating function.
In particular, for
\begin{eqnarray}
	X_\a = \Eint \sum_{\b\neq\a} \cT^{\a\b}_E \Sigma^{\a\b}_E X^{\a\b}_E 
		& = \frac{\kb}{h\sigma_\a}\Eint \sum_{\b\neq\a} \cT^{\a\b}_E
			(g^{\a\b}_E - g^{\b\a}_E)\\
		& = \frac{\kb J_\a}{\sigma_\a},\nonumber
\end{eqnarray}
the term proportional to $\Phi'(X_\a)$ vanishes and we are left with the bound
\begin{equation}\label{eq:apr:SCFBnd3}
	\chi_\a(s) \geq 2 J_\a\coth\bx\frac{\sigma_\a}{4\kb J_\a}\by
		\ln\bx\frac{\cosh[s/2+\sigma_\a/4\kb J_\a]}{
			\cosh[\sigma_\a/4\kb J_\a]}\by,
\end{equation}
which, as anticipated, involves only the mean particle current $J_\a$ and the partial rate of entropy production $\sigma_\a$. 

The constraint \eqref{eq:int:LDFBndQu} can now be deduced in two steps. 
First, we note that the right-hand side of the inequality \eqref{eq:apr:SCFBnd3} is monotonically decreasing in $\sigma_\a$ for $\sigma_\a\geq 0$, see \ref{app:SI}.
Therefore, it remains valid when the partial dissipation rate $\sigma_\a$ is replaced with the larger total dissipation rate $\sigma$.  
Upon introducing the dissipation factor 
\begin{equation}
	R_\a = \coth\bx\frac{\sigma}{4\kb J_\a}\by, 
\end{equation}
the resulting bound can be expressed in the compact form 
\begin{equation}\label{eq:apr:SCFBnd4}
	\chi_\a(s) \geq \hat{\chi}_\a(s) =  2J_\a R_\a\ln\bx\frac{R_\a\cosh[s/2] + \sinh[s/2]}{R_\a}\by. 
\end{equation}
Second, we recall that this result implies the upper bound 
\begin{equation}\label{eq:apr:BndLDFformal}
	I_\a(j) \leq \sup_s [js - \hat{\chi}_\a(s)]
\end{equation}
on the rate function $I_\a(j)$ of the particle current injected by the reservoir $\a$. 
To evaluate the Legendre-Fenchel transform, we note that 
\begin{eqnarray}
	\Bigl(js-\hat{\chi}_\a(s)\Bigr)' & = 
		j + J_\a R_\a \frac{1-R_\a + (1+R_\a)e^{s}}{1-R_\a - (1+R_\a)e^{s}},\\
	\Bigl(js-\hat{\chi}_\a(s)\Bigr)'' & = 
		2J_\a R_\a \frac{1-R_\a^2}{(1-R_\a -(1+R_\a)e^s)^2}\leq 0,
\end{eqnarray}
where the last inequality follows by observing that $J_\a R_\a\geq 0$ and $|R_\a|\geq 1$. 
Hence, the function $js-\hat{\chi}_\a(s)$ is concave and attains its global maximum for
\begin{equation}
	s = s^\ast = \ln\bx\frac{J_\a R_\a + j}{J_\a R_\a + J_\a}\by- \ln\bx\frac{J_\a R_\a - j}{J_\a R_\a -J_\a}\by
\end{equation}
Consequently, Eq.~\eqref{eq:apr:BndLDFformal} implies the bound 
\begin{eqnarray}
\label{eq:apr:BndLDFfinal}
\fl	I_\a(j) \leq \hat{I}^\a_\qu(j) & = js^\ast -\hat{\chi}_\a(s^\ast)\\
		& = (J_\a R_\a + j)\ln\bx\frac{J_\a R_\a +j}{J_\a R_\a + J_\a}\by
			+ (J_\a R_\a - j)\ln\bx\frac{J_\a R_\a -j}{J_\a R_\a - J_\a}\by, \nonumber
\end{eqnarray}
which holds for $-J_\a R_\a < j < J_\a R_\a$ and constitutes our main result. 

\subsection{Saturation}

To show that the bound \eqref{eq:apr:BndLDFfinal} can be saturated, at least in principle, we consider a two-terminal system, where both reservoirs have the same temperature $T$ and the chemical potentials $\mu_1 = E_0 + \kb T \cF/2$ and $\mu_2 = E_0 - \kb T \cF/2$.  
We further assume that each lead supports a single transversal mode at $E_0$ and that particles are transmitted only in a small energy window of width $w$ around $E_0$; that is, we set 
\begin{equation}
	\bT^{12}_E = \cT^{12}_E = \Pi_{E-E_0}(w),
\end{equation}
where $\Pi_E(w)$ denotes a boxcar function of height $1$ and width $w$ centered at $E=0$. 
Up to second-order corrections in $w$, the mean particle current injected by the reservoir $1$ and the total rate of entropy production are then given by 
\begin{eqnarray}
	J & = \frac{w}{h}(f^1_{E_0} - f^2_{E_0}) = \frac{w}{h}\tanh[\mathcal{F}/4],\\
	\sigma & = \frac{\kb w}{h}(g^{12}_{E_0} - g^{21}_{E_0})\ln[g^{12}_{E_0}/g^{21}_{E_0}] 
		=  \frac{\kb w}{h}\mathcal{F}\tanh[\mathcal{F}/4].
\end{eqnarray}
The dissipation factor thus becomes 
\begin{equation}
	R= \coth\bx\frac{\sigma}{4\kb J}\by = \coth[\cF/4]
\end{equation}
and the relation \eqref{eq:apr:SCFBnd4} takes the form 
\begin{equation}
	\chi_1(s) \geq \hat{\chi}_1(s) = \frac{w}{h}\ln\bx\frac{1+\cosh[\cF/2 +s]}{1+\cosh[\cF/2]}\by.
\end{equation}
Using the formula \eqref{eq:apr:LLFormT}, it is now straightforward to verify that this bound on the scaled cumulant generating function is indeed saturated up to higher-order corrections in $w$. 
The same is thus true for the corresponding bound \eqref{eq:apr:BndLDFfinal} on the rate function $I_1 (j)$, and the thermodynamic uncertainty relation \eqref{eq:int:TURQu}.

\subsection{Comparison with Classical Constraint}

We conclude the technical part of this paper by showing that the constraint \eqref{eq:apr:BndLDFfinal} is universally weaker than its classical counterpart 
\begin{equation}\label{eq:apr:ClThConstr}
	I_\a(j) \leq \hat{I}^\a_\cl = \frac{(j-J_\a)^2}{4\kb J^2_\a}\sigma.
\end{equation}
To this end, we define the function 
\begin{equation}\label{eq:apr:DefPsi}
	\phi(j) = \frac{\hat{I}_\cl(j) - \hat{I}_\qu(j)}{2},
\end{equation}
and consider its derivative 
\begin{equation}
	\phi'(j)  = \frac{j}{J}\arcoth[R]- \arcoth\bx \frac{JR}{j}\by\nonumber
\end{equation}
where we have used the definition $\arcoth[x] = \ln[(x+1)/(x-1)]/2$ of the inverse hyperbolic cotangent;
note that we drop the index $\a$ from Eq.~\eqref{eq:apr:DefPsi} onward for simplicity. 
Next, we observe that, for $x\geq 1$, the inequalities 
\begin{equation}\label{eq:apr:Coth}
	\arcoth[x] \geq \frac{1}{y}\arcoth\bx\frac{x}{y}\by \quad\mathrm{and}\quad
	\arcoth[x] \geq y\arcoth[x y]
\end{equation}
hold for any $|y|\leq 1$ and any $1 \leq |y|$, respectively, as we show in \ref{app:SI}. 
Upon taking $J,R >0$, we thus obtain the relations 
\begin{eqnarray}
	\phi'(j) \geq 0 &\quad\mathrm{for}\quad -RJ<j<-J &\quad\mathrm{and}\quad 0<j<J,\\
	\phi'(j) \leq 0 &\quad\mathrm{for}\quad -J<j<0   &\quad\mathrm{and}\quad J<j<RJ,
\end{eqnarray}
which imply that $\phi(j)$ has three local extrema in the open interval $-RJ < j < RJ$, a local minimum at $J=0$ and two local maxima at $j=-J$ and $j=J$. 
Since $\phi(-J) = \phi(J)=0$, it follows that $\phi(j)\leq 0$ for any  $-RJ < j < RJ$. 
The symmetry 
\begin{equation}
	\phi(j)|_{J\rightarrow -J, \; R\rightarrow -R} = \phi(-j)
\end{equation}
further shows that the same is true for $J,R<0$. 
Thus, we can conclude that
\begin{equation}
	\hat{I}_\cl (j) \leq \hat{I}_\qu (j)
\end{equation}
over the entire admissible range of the fluctuating current $j$.
Notably, the above inequality is asymptotically saturated in the low-dissipation limit, where $\sigma/J\rightarrow 0$ and $R\rightarrow \infty$. 
Specifically, we have 
\begin{equation}
	\hat{I}_\cl(j) = \frac{(j-J)^2}{JR} + O(R^{-3}) \quad\mathrm{and}\quad 
		\hat{I}_\qu(j) = \frac{(j-J)^2}{JR} + O(R^{-3}),
\end{equation}
and thus 
\begin{equation}
	\lim_{R\rightarrow\infty} \frac{\hat{I}_\cl(j)}{\hat{I}_\qu(j)} = 1. 
\end{equation}

\section{Example}\label{sec:exp}

\begin{figure}
\includegraphics[width=\textwidth]{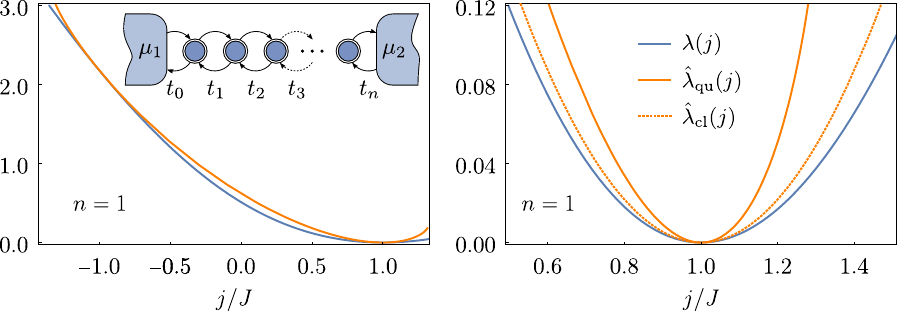}
\caption{
	\label{fig:exp:1}
	Current fluctuations in a single quantum dot. 
	\textbf{Left:} The sketch shows the considered system, which consists of $n$ quantum dots with nearest-neighbor tunneling, coupled to two reservoirs with
	the same temperature $T$ and chemical potentials $\mu_1 = E_0 + \kb T\cF/2$ and $\mu_2 = E_0 -\kb T\cF/2$. 
	The blue line shows the rate function $I(j)$ in units of $I_0 = \kb T/h$ as a function of the normalized fluctuating current $j/J$; 
	the orange line indicates the upper bound $\hat{I}_\qu(j)$. 
	\textbf{Right:} Behavior of $I(j)$ and $\hat{I}_\qu(j)$ around the mean current $J$. 
	For comparison, the dashed line shows the classical bound $\hat{I}_\cl(j)$. 
	For all plots, we have used the transmission function \eqref{eq:exp:TransFun} with $w=\kb T/4$ and $n=1$, 
	and set $\cF=4$ such that $\sigma/4\kb J =\cF/4 = 1$ and $R\simeq 1.31$.}
\end{figure}

\begin{figure}
\includegraphics[width=\textwidth]{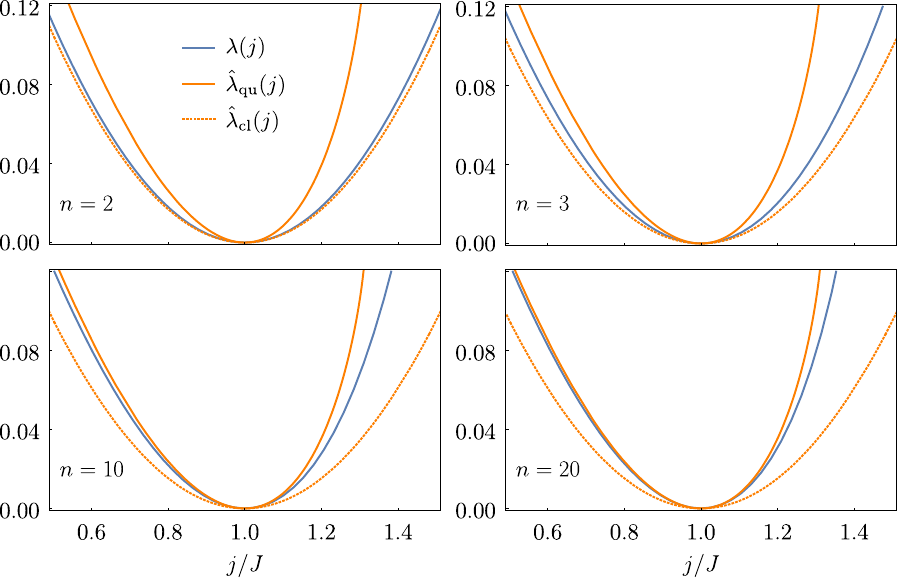}
\caption{
	\label{fig:exp:2}
	Quantum suppressed current fluctuations in chains of quantum dots. 
	The plots show the same quantities as the right panel of Fig.~\ref{fig:exp:1} for an increasing number $n$ of dots under otherwise identical conditions.
	The classical upper bound $\hat{I}_\cl(j)$ on $I(j)$ is violated for $n\geq 2$, while the quantum bound $\hat{I}_\qu(j)$ is almost 
	saturated for $n=20$.}
\end{figure}

As we have seen in the previous section, the thermodynamic constraint \eqref{eq:apr:BndLDFfinal} becomes tight in two-terminal settings with a narrow boxcar-shaped transmission function. 
Such a setup can be realized, for example, with chains of quantum dots forming a periodic potential landscape, which gives rise to a band-structured transmission spectrum \cite{whitney2015,ehrlich2021}. 
For discrete conductor of this type, the transmission function can be determined via the general formula \cite{whitney2015}
\begin{equation}\label{eq:exp:TransForm}
	\cT^{12}_E = \cT_E = 4\Gamma  \Bigl|\bra{n} [E-H_\mathrm{eff}]^{-1}\ket{1}\Bigr|^2.
\end{equation}
Here, $H_\mathrm{eff}$ denotes the effective single-particle Hamiltonian, which describes the tunneling dynamics between quantum dots and reservoirs, and $\ket{\ell}$ indicates the quantum state, where the particle is localized at the dot $\ell$. 
Furthermore, Eq.~\eqref{eq:exp:TransForm} assumes that particles enter and leave the conductor through the dots $1$ and $n$, which couple to the reservoirs $1$ and $2$, respectively, through the energy independent tunneling rate $\Gamma$.
For a chain of $n$ dots with nearest-neighbor coupling, like the system sketched in Fig.~\ref{fig:exp:1}, the effective Hamiltonian is given by 
\begin{equation}
	H_\mathrm{eff} = \sum_{\ell=1}^{n-1} t_\ell \Bigl(\ket{\ell}\bra{\ell+1} +\ket{\ell+1}\bra{\ell}\Bigr) -i\Gamma\ket{1}\bra{1} - i\Gamma\ket{n}\bra{n} +E_0,
\end{equation}
where we have taken the inter-dot tunnel couplings $t_\ell$ to be independent of energy, and $E_0$ is a global energy offset. 
As the length of the chain increases, the transmission function induced by this Hamiltonian generically develops an increasingly sharper peak around some resonant energy.
For finite chains, this peak is typically superimposed by coherent oscillations, causing the overall transmission function to deviate from its ideal shape. 
This effect can be suppressed, even for short chains, by fine tuning the tunneling energies. 
Here, we choose the specific transmission profile 
\begin{eqnarray}
	t_\ell & = t_{n-\ell} =
		 \frac{w}{2}\bx \sin\bx\frac{(2\ell+1)\pi}{2n}\by\sin\bx\frac{(2\ell -1)\pi}{2n}\by\by^{-\frac{1}{2}},\\
	i\Gamma & = t_0 = t_n,
\end{eqnarray}
whose corresponding transmission function 
\begin{equation}\label{eq:exp:TransFun}
	\cT_E = \frac{1}{1 + [2(E-E_0)/w]^{2n}}
\end{equation}
is free of oscillations and converges to a boxcar function of width $w$ in the limit $n\rightarrow\infty$ \cite{brandner2025c}.
Equipped with these prerequisites, we can determine the rate function $I_1(j) = I(j)$ by first calculating the corresponding scaled cumulant generating function via the formula \eqref{eq:apr:LLForm}, and then carrying out the Legendre-Fenchel transform \eqref{eq:apr:LFTransform} numerically. 
The mean current $J_1=J$ and the total rate of entropy production, which enter the thermodynamic constraints \eqref{eq:apr:BndLDFfinal} and \eqref{eq:apr:ClThConstr}, can be obtained directly from the formulas \eqref{eq:apr:DefCurrJ} and \eqref{eq:apr:TotEntProd}.
The results of these calculations are shown in Figs.~\ref{fig:exp:1} and \ref{fig:exp:2}. 
Notably, the classical bound $\hat{I}_\cl(j)$ still applies for a single quantum dot. 
For two and more dots, however, increasingly pronounced violations of this bound emerge and the rate function $I(j)$ approaches the weaker quantum bound $\hat{I}_\qu(j)$. 
That is, as the transmission function $\cT_E$ comes closer to its ideal boxcar shape, the probability to observe either typical or rare current fluctuations is more and more reduced below its classical minimum at a given total dissipation rate. 

This effect can be traced back to the same mechanism that has previously been identified as the cause of violations of the classical uncertainty relation \eqref{eq:int:TURCl} in coherent transport \cite{brandner2018,brandner2025c}. 
A quasi-classical transmission, which takes only the values $0$ and $1$, signifies that quantum fluctuations stemming from coherent superpositions between different scattering pathways have been eliminated. 
Thus, the only remaining source of current fluctuations are thermally induced variations in the occupation of individual scattering states. 
If the Fermi edges of the source and drain reservoirs are far above and far below the transmission window, respectively, the scattering states propagating in the direction of the bias are, on average, maximally occupied, while those propagating against the bias are essentially empty. 
In this situation, the Pauli exclusion principle, which is the physical origin of violations of the classical bounds \eqref{eq:int:TURCl} and \eqref{eq:apr:ClThConstr}, effectively makes the energetic cost of thermal fluctuations in the occupation of transmitting scattering states exponentially large in the applied bias. 
At the same time, the magnitude of the fluctuating current is ultimately limited by the number of available scattering states inside the transmission window, which explains why the bound \eqref{eq:apr:BndLDFfinal} applies only in a finite range of currents and, in contrast to its classical counterpart \eqref{eq:apr:ClThConstr}, is not symmetric around the mean current $J$. 
As a result, $J$ saturates to a finite value and the total rate of entropy production, which for the setting of Fig.~\ref{fig:exp:1} is given by $\sigma = (\mu_1 -\mu_2)J/T$, grows only linearly in the large-bias limit. 
This behavior is in sharp contrast to classical systems, where fluctuating currents are generally not subject to fundamental limits and Fermi distributions are replaced with Boltzmann distributions, which impose much less stringent constraints on fluctuations. 
Finally, the picture outlined above also clarifies why no violations of the classical uncertainty relation \eqref{eq:int:TURCl} can be observed in transport settings with Bosonic particles, or quasi-particles \cite{saryal2019}. 

\section{Perspectives}\label{sec:per}

We have derived a universal thermodynamic constraint on the rate functions of fluctuating particle currents in coherent multi-terminal conductors. 
This constraint holds for any sample geometry and potential landscape, and under arbitrary temperature and chemical potential biases. 
It implies the recently discovered generalized thermodynamic uncertainty relation \eqref{eq:int:TURQu} and remains robust against moderate dephasing and inelastic scattering, provided that these effects can be modeled by virtual terminals that account for incoherent energy exchange between particles, or between particles and additional degrees of freedom, at a mean-field level.

The results summarized above follow from a symmetry relation between coherent transmission coefficients, which in general requires the microscopic dynamics inside the conductor to be invariant under time reversal, at least for settings with more than two terminals. 
At this point, our analysis does not cover setups where this symmetry is broken, for instance due to an external magnetic field. 
Notably, the uncertainty relation \eqref{eq:int:TURQu} can be adapted to such situations by introducing a numerical factor. 
Specifically, the bound 
\begin{equation}\label{eq:per:GenTUR}
	\frac{S}{J}\sinh\bx\frac{\sigma}{\psi_0\kb J}\by \geq 1 \quad\mathrm{with}\quad \psi_0\simeq 0.85246 \geq 17/20
\end{equation}
holds for any particle current in a multi-terminal coherent conductor, regardless of how the underlying micro-dynamics behave under time reversal \cite{brandner2025c}. 
It is thus tempting to speculate that the thermodynamic constraint \eqref{eq:int:LDFBndQu} could be extended to systems with broken time-reversal symmetry by inserting $\psi_0$ into the dissipation factor \eqref{eq:int:DissFact} such that the bound \eqref{eq:per:GenTUR} is recovered for the second cumulant.
Whether or not this conjecture can be corroborated remains to be seen. 
At the same time, we leave it to future research to investigate whether constraints similar to the one derived here might apply also to thermodynamic quantities other than particle currents. 
The most natural candidates for such quantities would be heat and energy currents, for which none of the bounds \eqref{eq:int:TURQu}, \eqref{eq:int:LDFBndQu} or \eqref{eq:per:GenTUR} can apply due to a mismatch of physical dimensions. 
For these quantities, it is, to the best of our knowledge, not yet clear under what conditions strong violations of the classical bounds \eqref{eq:int:LDFBndCl} and \eqref{eq:int:TURCl}, which cover all types of currents, can be generated, and thus how less restrictive quantum bounds should be constructed.
Finally, it would be most interesting to explore how the ideas developed in this article can be extended to more complex dynamical scenarios involving, for example, strongly interacting particles or superconducting reservoirs \cite{ohnmacht2025,arrachea2025}. 
 
\section*{Acknowledgments}

This work was supported by the Medical Research Council (Grants No. MR/S034714/1 and MR/Y003845/1) and the Engineering and Physical Sciences Research Council (Grant No. EP/V031201/1).
K.S. was supported by the JSPS KAKENHI Grant No. JP23K25796.

\section*{Data Availability Statement}

All data that support the findings of this study are included within the article. 

\section*{References}

\bibliographystyle{utphys}

\begin{thebibliography}{10}

\bibitem{fujiwara2008}
A.~Fujiwara, K.~Nishiguchi, and Y.~Ono, ``Nanoampere charge pump by
  single-electron ratchet using silicon nanowire metal-oxide-semiconductor
  field-effect transistor,'' \href{http://dx.doi.org/10.1063/1.2837544}{{\em
  Appl. Phys. Lett.} {\bfseries 92} (2008) 042102}.

\bibitem{bitton2017}
O.~Bitton, D.~B. Gutman, R.~Berkovits, and A.~Frydman, ``Multiple periodicity
  in a nanoparticle-based single-electron transistor,''
  \href{http://dx.doi.org/10.1038/s41467-017-00442-6}{{\em Nat. Commun.}
  {\bfseries 8} (2017) 402}.

\bibitem{asgari2021}
M.~Asgari, D.~Coquillat, G.~Menichetti, V.~Zannier, N.~Diakonova, W.~Knap,
  L.~Sorba, L.~Viti, and M.~S. Vitiello, ``Quantum-{{Dot Single-Electron
  Transistors}} as {{Thermoelectric Quantum Detectors}} at {{Terahertz
  Frequencies}},'' \href{http://dx.doi.org/10.1021/acs.nanolett.1c02022}{{\em
  Nano Lett.} {\bfseries 21} (2021) 8587--8594}.

\bibitem{ji2003}
Y.~Ji, Y.~Chung, D.~Sprinzak, M.~Heiblum, D.~Mahalu, and H.~Shtrikman, ``An
  electronic {{Mach}}--{{Zehnder}} interferometer,''
  \href{http://dx.doi.org/10.1038/nature01503}{{\em Nature} {\bfseries 422}
  (2003) 415--418}.

\bibitem{jo2021}
M.~Jo, P.~Brasseur, A.~Assouline, G.~Fleury, H.-S. Sim, K.~Watanabe,
  T.~Taniguchi, W.~Dumnernpanich, P.~Roche, D.~C. Glattli, N.~Kumada, F.~D.
  Parmentier, and P.~Roulleau, ``Quantum {{Hall Valley Splitters}} and a
  {{Tunable Mach-Zehnder Interferometer}} in {{Graphene}},''
  \href{http://dx.doi.org/10.1103/PhysRevLett.126.146803}{{\em Phys. Rev.
  Lett.} {\bfseries 126} (2021) 146803}.

\bibitem{chakraborti2025}
H.~Chakraborti, L.~Pugliese, A.~Assouline, K.~Watanabe, T.~Taniguchi,
  N.~Kumada, D.~C. Glattli, M.~Jo, H.-S. Sim, and P.~Roulleau, ``Electron
  collision in a two-path graphene interferometer,''
  \href{http://dx.doi.org/10.1126/science.adn4622}{{\em Science} {\bfseries
  388} (2025) 492--496}.

\bibitem{prance2009}
J.~R. Prance, C.~G. Smith, J.~P. Griffiths, S.~J. Chorley, D.~Anderson,
  G.~A.~C. Jones, I.~Farrer, and D.~A. Ritchie, ``Electronic {{Refrigeration}}
  of a {{Two-Dimensional Electron Gas}},''
  \href{http://dx.doi.org/10.1103/PhysRevLett.102.146602}{{\em Phys. Rev.
  Lett.} {\bfseries 102} (2009) 146602}.

\bibitem{brantut2013}
J.-P. Brantut, C.~Grenier, J.~Meineke, D.~Stadler, S.~Krinner, C.~Kollath,
  T.~Esslinger, and A.~Georges, ``A {{Thermoelectric Heat Engine}} with
  {{Ultracold Atoms}},'' \href{http://dx.doi.org/10.1126/science.1242308}{{\em
  Science} {\bfseries 342} (2013) 713--715}.

\bibitem{josefsson2018}
M.~Josefsson, A.~Svilans, A.~M. Burke, E.~A. Hoffmann, S.~Fahlvik,
  C.~Thelander, M.~Leijnse, and H.~Linke, ``A quantum-dot heat engine operating
  close to the thermodynamic efficiency limits,''
  \href{http://dx.doi.org/10.1038/s41565-018-0200-5}{{\em Nat. Nanotech.}
  {\bfseries 13} (2018) 920--924}.

\bibitem{seifert2012}
U.~Seifert, ``Stochastic thermodynamics, fluctuation theorems and molecular
  machines,'' \href{http://dx.doi.org/10.1088/0034-4885/75/12/126001}{{\em Rep.
  Prog. Phys.} {\bfseries 75} (2012) 126001}.

\bibitem{benenti2017}
G.~Benenti, G.~Casati, K.~Saito, and R.~S. Whitney, ``Fundamental aspects of
  steady-state conversion of heat to work at the nanoscale,''
  \href{http://dx.doi.org/10.1016/j.physrep.2017.05.008}{{\em Physics Reports}
  {\bfseries 694} (2017) 1--124}.

\bibitem{bauerle2018}
C.~B{\"a}uerle, D.~Christian~Glattli, T.~Meunier, F.~Portier, P.~Roche,
  P.~Roulleau, S.~Takada, and X.~Waintal, ``Coherent control of single
  electrons: A review of current progress,''
  \href{http://dx.doi.org/10.1088/1361-6633/aaa98a}{{\em Rep. Prog. Phys.}
  {\bfseries 81} (2018) 056503}.

\bibitem{andrieux2006}
D.~Andrieux and P.~Gaspard, ``Fluctuation theorem for transport in mesoscopic
  systems,'' \href{http://dx.doi.org/10.1088/1742-5468/2006/01/P01011}{{\em J.
  Stat. Mech.} {\bfseries 2006} (2006) P01011}.

\bibitem{andrieux2007}
D.~Andrieux and P.~Gaspard, ``A fluctuation theorem for currents and non-linear
  response coefficients,''
  \href{http://dx.doi.org/10.1088/1742-5468/2007/02/P02006}{{\em J. Stat.
  Mech.} {\bfseries 2007} (2007) P02006}.

\bibitem{saito2008}
K.~Saito and Y.~Utsumi, ``Symmetry in full counting statistics, fluctuation
  theorem, and relations among nonlinear transport coefficients in the presence
  of a magnetic field,''
  \href{http://dx.doi.org/10.1103/PhysRevB.78.115429}{{\em Phys. Rev. B}
  {\bfseries 78} (2008) 115429}.

\bibitem{nakamura2010}
S.~Nakamura, Y.~Yamauchi, M.~Hashisaka, K.~Chida, K.~Kobayashi, T.~Ono,
  R.~Leturcq, K.~Ensslin, K.~Saito, Y.~Utsumi, and A.~C. Gossard,
  ``Nonequilibrium {{Fluctuation Relations}} in a {{Quantum Coherent
  Conductor}},'' \href{http://dx.doi.org/10.1103/PhysRevLett.104.080602}{{\em
  Phys. Rev. Lett.} {\bfseries 104} (2010) 080602}.

\bibitem{sanchez2010}
R.~S{\'a}nchez, R.~L{\'o}pez, D.~S{\'a}nchez, and M.~B{\"u}ttiker, ``Mesoscopic
  {{Coulomb Drag}}, {{Broken Detailed Balance}}, and {{Fluctuation
  Relations}},'' \href{http://dx.doi.org/10.1103/PhysRevLett.104.076801}{{\em
  Phys. Rev. Lett.} {\bfseries 104} (2010) 076801}.

\bibitem{hussein2014}
R.~Hussein and S.~Kohler, ``Quantum transport, master equations, and exchange
  fluctuations,'' \href{http://dx.doi.org/10.1103/PhysRevB.89.205424}{{\em
  Phys. Rev. B} {\bfseries 89} (2014) 205424}.

\bibitem{gaspard2015}
P.~Gaspard, ``Scattering approach to the thermodynamics of quantum transport,''
  \href{http://dx.doi.org/10.1088/1367-2630/17/4/045001}{{\em New J. Phys.}
  {\bfseries 17} (2015) 045001}.

\bibitem{potanina2021}
E.~Potanina, C.~Flindt, M.~Moskalets, and K.~Brandner, ``Thermodynamic bounds
  on coherent transport in periodically driven conductors,''
  \href{http://dx.doi.org/10.1103/PhysRevX.11.021013}{{\em Phys. Rev. X}
  {\bfseries 11} (2021) 021013}.

\bibitem{eriksson2021}
J.~Eriksson, M.~Acciai, L.~Tesser, and J.~Splettstoesser, ``General {{Bounds}}
  on {{Electronic Shot Noise}} in the {{Absence}} of {{Currents}},''
  \href{http://dx.doi.org/10.1103/PhysRevLett.127.136801}{{\em Phys. Rev.
  Lett.} {\bfseries 127} (2021) 136801}.

\bibitem{tesser2023}
L.~Tesser, M.~Acciai, C.~Sp{\aa}nsl{\"a}tt, J.~Monsel, and J.~Splettstoesser,
  ``Charge, spin, and heat shot noises in the absence of average currents:
  {{Conditions}} on bounds at zero and finite frequencies,''
  \href{http://dx.doi.org/10.1103/PhysRevB.107.075409}{{\em Phys. Rev. B}
  {\bfseries 107} (2023) 075409}.

\bibitem{tesser2024}
L.~Tesser and J.~Splettstoesser, ``Out-of-{{Equilibrium Fluctuation-Dissipation
  Bounds}},'' \href{http://dx.doi.org/10.1103/PhysRevLett.132.186304}{{\em
  Phys. Rev. Lett.} {\bfseries 132} (2024) 186304}.

\bibitem{acciai2024}
M.~Acciai, L.~Tesser, J.~Eriksson, R.~S{\'a}nchez, R.~S. Whitney, and
  J.~Splettstoesser, ``Constraints between entropy production and its
  fluctuations in nonthermal engines,''
  \href{http://dx.doi.org/10.1103/PhysRevB.109.075405}{{\em Phys. Rev. B}
  {\bfseries 109} (2024) 075405}.

\bibitem{palmqvist2024}
D.~Palmqvist, L.~Tesser, and J.~Splettstoesser, ``Kinetic uncertainty relations
  for quantum transport,''
  \href{http://arxiv.org/abs/2410.10793}{arXiv:2410.10793 (2024)}.

\bibitem{palmqvist2025}
D.~Palmqvist, L.~Tesser, and J.~Splettstoesser, ``Combining kinetic and
  thermodynamic uncertainty relations in quantum transport,''
  \href{http://arxiv.org/abs/2504.04980}{arXiv:2504.04980 (2025)}.

\bibitem{blasi2025}
G.~Blasi, R.~R. Rodr{\'i}guez, M.~Moskalets, R.~L{\'o}pez, and G.~Haack,
  ``Quantum {{Kinetic Uncertainty Relations}} in {{Mesoscopic Conductors}} at
  {{Strong Coupling}},''
  \href{http://arxiv.org/abs/2505.13200}{arXiv:2505.13200 (2025)}.

\bibitem{brandner2025c}
K.~Brandner and K.~Saito, ``Thermodynamic {{Uncertainty Relations}} for
  {{Coherent Transport}},''
  \href{http://arxiv.org/abs/2502.07917}{arXiv:2502.07917 (2025)}.

\bibitem{seifert2019}
U.~Seifert, ``From {{Stochastic Thermodynamics}} to {{Thermodynamic
  Inference}},''
  \href{http://dx.doi.org/10.1146/annurev-conmatphys-031218-013554}{{\em Annu.
  Rev. Condens. Matter Phys.} {\bfseries 10} (2019) 171--192}.

\bibitem{horowitz2020}
J.~M. Horowitz and T.~R. Gingrich, ``Thermodynamic uncertainty relations
  constrain non-equilibrium fluctuations,''
  \href{http://dx.doi.org/10.1038/s41567-019-0702-6}{{\em Nat. Phys.}
  {\bfseries 16} (2020) 15--20}.

\bibitem{pietzonka2018}
P.~Pietzonka and U.~Seifert, ``Universal {{Trade-Off}} between {{Power}},
  {{Efficiency}}, and {{Constancy}} in {{Steady-State Heat Engines}},''
  \href{http://dx.doi.org/10.1103/PhysRevLett.120.190602}{{\em Phys. Rev.
  Lett.} {\bfseries 120} (2018) 190602}.

\bibitem{koyuk2019}
T.~Koyuk and U.~Seifert, ``Operationally {{Accessible Bounds}} on
  {{Fluctuations}} and {{Entropy Production}} in {{Periodically Driven
  Systems}},'' \href{http://dx.doi.org/10.1103/PhysRevLett.122.230601}{{\em
  Phys. Rev. Lett.} {\bfseries 122} (2019) 230601}.

\bibitem{touchette2009a}
H.~Touchette, ``The large deviation approach to statistical mechanics,''
  \href{http://dx.doi.org/10.1016/j.physrep.2009.05.002}{{\em Phys. Rep.}
  {\bfseries 478} (2009) 1--69}.

\bibitem{jack2020}
R.~L. Jack, ``Ergodicity and large deviations in physical systems with
  stochastic dynamics,''
  \href{http://dx.doi.org/10.1140/epjb/e2020-100605-3}{{\em Eur. Phys. J. B}
  {\bfseries 93} (2020) 74}.

\bibitem{burenev2025}
I.~N. Burenev, D.~W.~H. Cloete, V.~Kharbanda, and H.~Touchette, ``An
  introduction to large deviations with applications in physics,''
  \href{http://arxiv.org/abs/2503.16015}{arXiv:2503.16015 (2025)}.

\bibitem{pietzonka2016}
P.~Pietzonka, A.~C. Barato, and U.~Seifert, ``Universal bounds on current
  fluctuations,'' \href{http://dx.doi.org/10.1103/PhysRevE.93.052145}{{\em
  Phys. Rev. E} {\bfseries 93} (2016) 052145}.

\bibitem{gingrich2016}
T.~R. Gingrich, J.~M. Horowitz, N.~Perunov, and J.~L. England, ``Dissipation
  {{Bounds All Steady-State Current Fluctuations}},''
  \href{http://dx.doi.org/10.1103/PhysRevLett.116.120601}{{\em Phys. Rev.
  Lett.} {\bfseries 116} (2016) 120601}.

\bibitem{macieszczak2018}
K.~Macieszczak, K.~Brandner, and J.~P. Garrahan, ``Unified {{Thermodynamic
  Uncertainty Relations}} in {{Linear Response}},''
  \href{http://dx.doi.org/10.1103/PhysRevLett.121.130601}{{\em Phys. Rev.
  Lett.} {\bfseries 121} (2018) 130601}.

\bibitem{carollo2019}
F.~Carollo, R.~L. Jack, and J.~P. Garrahan, ``Unraveling the {{Large Deviation
  Statistics}} of {{Markovian Open Quantum Systems}},''
  \href{http://dx.doi.org/10.1103/PhysRevLett.122.130605}{{\em Phys. Rev.
  Lett.} {\bfseries 122} (2019) 130605}.

\bibitem{guarnieri2019}
G.~Guarnieri, G.~T. Landi, S.~R. Clark, and J.~Goold, ``Thermodynamics of
  precision in quantum nonequilibrium steady states,''
  \href{http://dx.doi.org/10.1103/PhysRevResearch.1.033021}{{\em Phys. Rev.
  Research} {\bfseries 1} (2019) 033021}.

\bibitem{hasegawa2020}
Y.~Hasegawa, ``Quantum {{Thermodynamic Uncertainty Relation}} for {{Continuous
  Measurement}},'' \href{http://dx.doi.org/10.1103/PhysRevLett.125.050601}{{\em
  Phys. Rev. Lett.} {\bfseries 125} (2020) 050601}.

\bibitem{hasegawa2021}
Y.~Hasegawa, ``Thermodynamic {{Uncertainty Relation}} for {{General Open
  Quantum Systems}},''
  \href{http://dx.doi.org/10.1103/PhysRevLett.126.010602}{{\em Phys. Rev.
  Lett.} {\bfseries 126} (2021) 010602}.

\bibitem{rignon-bret2021}
A.~{Rignon-Bret}, G.~Guarnieri, J.~Goold, and M.~T. Mitchison, ``Thermodynamics
  of precision in quantum nanomachines,''
  \href{http://dx.doi.org/10.1103/PhysRevE.103.012133}{{\em Phys. Rev. E}
  {\bfseries 103} (2021) 012133}.

\bibitem{menczel2021}
P.~Menczel, E.~Loisa, K.~Brandner, and C.~Flindt, ``Thermodynamic uncertainty
  relations for coherently driven open quantum systems,''
  \href{http://dx.doi.org/10.1088/1751-8121/ac0c8f}{{\em J. Phys. A: Math.
  Theor.} {\bfseries 54} (2021) 314002}.

\bibitem{vanvu2022}
T.~Van~Vu and K.~Saito, ``Thermodynamics of {{Precision}} in {{Markovian Open
  Quantum Dynamics}},''
  \href{http://dx.doi.org/10.1103/PhysRevLett.128.140602}{{\em Phys. Rev.
  Lett.} {\bfseries 128} (2022) 140602}.

\bibitem{prech2023}
K.~Prech, P.~Johansson, E.~Nyholm, G.~T. Landi, C.~Verdozzi, P.~Samuelsson, and
  P.~P. Potts, ``Entanglement and thermokinetic uncertainty relations in
  coherent mesoscopic transport,''
  \href{http://dx.doi.org/10.1103/PhysRevResearch.5.023155}{{\em Phys. Rev.
  Research} {\bfseries 5} (2023) 023155}.

\bibitem{vu2024}
T.~V. Vu, ``Fundamental bounds on precision and response for quantum trajectory
  observables,'' 
  \href{http://arxiv.org/abs/2411.19546}{arXiv:2411.19546 (2024)}.

\bibitem{moreira2024}
S.~V. Moreira, M.~Radaelli, A.~Candeloro, F.~C. Binder, and M.~T. Mitchison,
  ``Precision bounds for multiple currents in open quantum systems,'' 
  \href{http://arxiv.org/abs/2411.09088}{arXiv:2411.09088 (2024)}.

\bibitem{kwon2024}
E.~Kwon and J.~S. Lee, ``A unified framework for classical and quantum
  uncertainty relations using stochastic representations,''
  \href{http://arxiv.org/abs/2412.04988}{arXiv:2412.04988 (2024)}.

\bibitem{vanwees1988}
B.~J. Van~Wees, H.~Van~Houten, C.~W.~J. Beenakker, J.~G. Williamson, L.~P.
  Kouwenhoven, D.~Van Der~Marel, and C.~T. Foxon, ``Quantized conductance of
  point contacts in a two-dimensional electron gas,''
  \href{http://dx.doi.org/10.1103/PhysRevLett.60.848}{{\em Phys. Rev. Lett.}
  {\bfseries 60} (1988) 848--850}.

\bibitem{wharam1988}
D.~A. Wharam, T.~J. Thornton, R.~Newbury, M.~Pepper, H.~Ahmed, J.~E.~F. Frost,
  D.~G. Hasko, D.~C. Peacock, D.~A. Ritchie, and G.~A.~C. Jones,
  ``One-dimensional transport and the quantisation of the ballistic
  resistance,'' \href{http://dx.doi.org/10.1088/0022-3719/21/8/002}{{\em J.
  Phys. C: Solid State Phys.} {\bfseries 21} (1988) L209--L214}.

\bibitem{schwab2000}
K.~Schwab, E.~A. Henriksen, J.~M. Worlock, and M.~L. Roukes, ``Measurement of
  the quantum of thermal conductance,''
  \href{http://dx.doi.org/10.1038/35010065}{{\em Nature} {\bfseries 404} (2000)
  974--977}.

\bibitem{matthews2014}
J.~Matthews, F.~Battista, D.~S{\'a}nchez, P.~Samuelsson, and H.~Linke,
  ``Experimental verification of reciprocity relations in quantum
  thermoelectric transport,''
  \href{http://dx.doi.org/10.1103/PhysRevB.90.165428}{{\em Phys. Rev. B}
  {\bfseries 90} (2014) 165428}.

\bibitem{krans1995}
J.~M. Krans, J.~M. Van~Ruitenbeek, V.~V. Fisun, I.~K. Yanson, and L.~J.
  De~Jongh, ``The signature of conductance quantization in metallic point
  contacts,'' \href{http://dx.doi.org/10.1038/375767a0}{{\em Nature} {\bfseries
  375} (1995) 767--769}.

\bibitem{vandenbrom1999}
H.~E. Van Den~Brom and J.~M. Van~Ruitenbeek, ``Quantum {{Suppression}} of
  {{Shot Noise}} in {{Atom-Size Metallic Contacts}},''
  \href{http://dx.doi.org/10.1103/PhysRevLett.82.1526}{{\em Phys. Rev. Lett.}
  {\bfseries 82} (1999) 1526--1529}.

\bibitem{cui2017}
L.~Cui, W.~Jeong, S.~Hur, M.~Matt, J.~C. Kl{\"o}ckner, F.~Pauly, P.~Nielaba,
  J.~C. Cuevas, E.~Meyhofer, and P.~Reddy, ``Quantized thermal transport in
  single-atom junctions,''
  \href{http://dx.doi.org/10.1126/science.aam6622}{{\em Science} {\bfseries
  355} (2017) 1192--1195}.

\bibitem{lumbroso2018}
O.~S. Lumbroso, L.~Simine, A.~Nitzan, D.~Segal, and O.~Tal, ``Electronic noise
  due to temperature differences in atomic-scale junctions,''
  \href{http://dx.doi.org/10.1038/s41586-018-0592-2}{{\em Nature} {\bfseries
  562} (2018) 240--244}.

\bibitem{brantut2012}
J.-P. Brantut, J.~Meineke, D.~Stadler, S.~Krinner, and T.~Esslinger,
  ``Conduction of {{Ultracold Fermions Through}} a {{Mesoscopic Channel}},''
  \href{http://dx.doi.org/10.1126/science.1223175}{{\em Science} {\bfseries
  337} (2012) 1069--1071}.

\bibitem{krinner2015}
S.~Krinner, D.~Stadler, D.~Husmann, J.-P. Brantut, and T.~Esslinger,
  ``Observation of quantized conductance in neutral matter,''
  \href{http://dx.doi.org/10.1038/nature14049}{{\em Nature} {\bfseries 517}
  no.~7532, (2015) 64--67}.

\bibitem{lebrat2018}
M.~Lebrat, P.~Gri{\v s}ins, D.~Husmann, S.~H{\"a}usler, L.~Corman,
  T.~Giamarchi, J.-P. Brantut, and T.~Esslinger, ``Band and {{Correlated
  Insulators}} of {{Cold Fermions}} in a {{Mesoscopic Lattice}},''
  \href{http://dx.doi.org/10.1103/PhysRevX.8.011053}{{\em Phys. Rev. X}
  {\bfseries 8} (2018) 011053}.

\bibitem{sivan1986}
U.~Sivan and Y.~Imry, ``Multichannel {{Landauer}} formula for thermoelectric
  transport with application to thermopower near the mobility edge,''
  \href{http://dx.doi.org/10.1103/PhysRevB.33.551}{{\em Phys. Rev. B}
  {\bfseries 33} (1986) 551--558}.

\bibitem{buttiker1992}
M.~B{\"u}ttiker, ``Scattering theory of current and intensity noise
  correlations in conductors and wave guides,''
  \href{http://dx.doi.org/10.1103/PhysRevB.46.12485}{{\em Phys. Rev. B}
  {\bfseries 46} (1992) 12485--12507}.

\bibitem{brandner2018}
K.~Brandner, T.~Hanazato, and K.~Saito, ``Thermodynamic {{Bounds}} on
  {{Precision}} in {{Ballistic Multiterminal Transport}},''
  \href{http://dx.doi.org/10.1103/PhysRevLett.120.090601}{{\em Phys. Rev.
  Lett.} {\bfseries 120} (2018) 090601}.

\bibitem{saryal2019}
S.~Saryal, H.~M. Friedman, D.~Segal, and B.~K. Agarwalla, ``Thermodynamic
  uncertainty relation in thermal transport,''
  \href{http://dx.doi.org/10.1103/PhysRevE.100.042101}{{\em Phys. Rev. E}
  {\bfseries 100} (2019) 042101}.

\bibitem{ehrlich2021}
T.~Ehrlich and G.~Schaller, ``Broadband frequency filters with quantum dot
  chains,'' \href{http://dx.doi.org/10.1103/PhysRevB.104.045424}{{\em Phys.
  Rev. B} {\bfseries 104} (2021) 045424}.

\bibitem{gerry2022}
M.~Gerry and D.~Segal, ``Absence and recovery of cost-precision tradeoff
  relations in quantum transport,''
  \href{http://dx.doi.org/10.1103/PhysRevB.105.155401}{{\em Phys. Rev. B}
  {\bfseries 105} (2022) 155401}.

\bibitem{timpanaro2025}
A.~M. Timpanaro, G.~Guarnieri, and G.~T. Landi, ``Quantum thermoelectric
  transmission functions with minimal current fluctuations,'' 
  \href{http://arxiv.org/abs/2106.10205}{arXiv:2106.10205 (2025)}.

\bibitem{buttiker1988}
M.~B{\"u}ttiker, ``Coherent and sequential tunneling in series barriers,''
  \href{http://dx.doi.org/10.1147/rd.321.0063}{{\em IBM J. Res. \& Dev.}
  {\bfseries 32} (1988) 63--75}.

\bibitem{nenciu2007}
G.~Nenciu, ``Independent electron model for open quantum systems:
  {{Landauer-B{\"u}ttiker}} formula and strict positivity of the entropy
  production,'' \href{http://dx.doi.org/10.1063/1.2712418}{{\em J. Math. Phys.}
  {\bfseries 48} (2007) 033302}.

\bibitem{levitov1993}
L.~S. Levitov and G.~B. Lesovik, ``Charge distribution in quantum shot noise,''
  \href{http://jetpletters.ru/ps/1186/}{
  {\em J. Exp. Theor. Phys. Lett.} {\bfseries 58} (1993) 230.}

\bibitem{lesovik2011}
G.~B. Lesovik and I.~A. Sadovskyy, ``Scattering matrix approach to the
  description of quantum electron transport,''
  \href{http://dx.doi.org/10.3367/UFNe.0181.201110b.1041}{{\em Phys.-Usp.}
  {\bfseries 54} (2011) 1007--1059}.

\bibitem{hansen1982}
F.~Hansen and G.~Kj{\ae}rg{\aa}rd~Pedersen, ``Jensen's inequality for operators
  and {{L{\"o}wner}}'s theorem,''
  \href{http://dx.doi.org/10.1007/bf01450679}{{\em Math. Ann.} {\bfseries 258}
  (1982) 229--241}.

\bibitem{whitney2015}
R.~S. Whitney, ``Finding the quantum thermoelectric with maximal efficiency and
  minimal entropy production at given power output,''
  \href{http://dx.doi.org/10.1103/PhysRevB.91.115425}{{\em Phys. Rev. B}
  {\bfseries 91} (2015) 115425}.

\bibitem{ohnmacht2025}
D.~C. Ohnmacht, J.~C. Cuevas, W.~Belzig, R.~L{\'o}pez, J.~S. Lim, and K.~W.
  Kim, ``Thermodynamic uncertainty relations in superconducting junctions,''
  \href{http://dx.doi.org/10.1103/PhysRevResearch.7.L012075}{{\em Phys. Rev.
  Res.} {\bfseries 7} (2025) L012075}.

\bibitem{arrachea2025}
L.~Arrachea, A.~Braggio, P.~Burset, E.~J.~H. Lee, A.~L. Yeyati, and
  R.~S{\'a}nchez, ``Thermoelectric processes of quantum normal-superconductor
  interfaces,''
  \href{http://arxiv.org/abs/2505.07426}{arXiv:2505.07426 (2025)}.

\end{thebibliography}

\providecommand{\href}[2]{#2}\begingroup\raggedright\endgroup

\newpage
\appendix
\section{Lemmas}\label{app:SI}

\newcounter{lemma}
\newcommand{\uselemma}{\stepcounter{lemma}\thelemma}

\noindent
\textbf{Lemma \uselemma.} For any real $s$, the function
\begin{equation}
	\phi(x) = x\coth\bx\frac{1}{x}\by\ln\bx\frac{\cosh[s+1/x]}{\cosh[1/x]}\by
\end{equation}
is convex over the entire real axis.\\

\noindent
\emph{Proof.}
We first note that the definition above is equivalent to 
\begin{equation}
	\phi(x) = x\coth\bx\frac{1}{x}\by \ln\bx [1+u^2]^\frac{1}{2} + u\tanh\bx\frac{1}{x}\by\by
\end{equation}
with $u=\sinh[s]$. 
Next, we calculate the second derivative of $\phi(x)$, which is given by 
\begin{equation}
	\phi''(x)=\frac{\cosh[1/x]}{x^3\sinh^3[1/x]}\biggl(\rho[w^2] + \frac{u^2(1-v^2)^2}{w^2}\biggr)
\end{equation}
with $v=\tanh[1/x]$, $w=[1+u^2]^\frac{1}{2}+uv$ and $\rho(x)= \ln[x] - 1 + 1/x$. 
This compact expression shows that $\phi''(x)\geq 0$ if $\rho(x)\geq 0$ for any $x\geq 0$. 
The latter statement can now be easily verified by calculating the derivative $\rho'(x) = (x-1)/x^2$, which is strictly negative for $x<1$ and strictly positive for $x>1$, implying that $\rho(x)$ has a global minimum at $x=1$. 
Since $\rho(1) = 0$, it follows that $\rho(x)\geq 0$ for any $x\geq 0$.\\
 
\noindent
\textbf{Lemma \uselemma.} For any real $J$ and $s$, the function
\begin{equation}
	\phi(\sigma) = J \coth\bx\frac{\sigma}{J}\by\ln\bx\frac{\cosh[s+\sigma/J]}{\cosh[\sigma/J]}\by
\end{equation}
is monotonically decreasing over the positive real half-axis.\\

\noindent
\emph{Proof.} We first consider the auxiliary function 
\begin{equation}
	\rho(x) = x\ln\bx [1+u^2]^\frac{1}{2} + \frac{J u}{x}\by
\end{equation}
for $x\geq 0$, where $u$ and $J$ are real parameters. 
The first and the second derivative of this function are given by 
\begin{eqnarray}
	\rho'(x) &= - \frac{J u}{Ju + x [1+u^2]^\frac{1}{2}} + \ln\bx [1+u^2]^\frac{1}{2} + \frac{J u}{x}\by,\\
	\rho''(x)&= - \frac{J^2 u^2}{x}\Bigl[Ju + x [1+u^2]^\frac{1}{2}\Bigr]^{-2}.
\end{eqnarray}
The fact that $\rho''(x)<0$ for any finite $x\geq 0$ shows that $\rho'(x)$ is monotonically decreasing.
Since $\lim_{x\rightarrow\infty}\rho'(x) = \ln[1+u^2]/2\geq 0$, it follows that $\rho'(x)\geq 0$ for any non-negative $x$. 
Hence, $\rho(x)$ is monotonically increasing. 
Upon setting $u=\sinh[s]$, we have 
\begin{equation}
	(\rho\circ \psi)(\sigma) = \phi(\sigma)
\end{equation}
with $\psi(\sigma) = J\coth[\sigma/J]$. 
Since $\psi(\sigma)$ is non-negative and monotonically decreasing over the positive real half-axis, $\phi(\sigma)$ must be monotonically decreasing over the same domain.\\

\noindent
\textbf{Lemma \uselemma.} For any $x\geq 1$, the inequalities 
\begin{equation}
	\arcoth[x] \geq \frac{1}{y}\arcoth\bx\frac{x}{y}\by\quad\mathrm{and}\quad
	\arcoth[x] \geq y\arcoth[x y]
\end{equation}
hold for any $|y| \leq 1$ and any $1\leq |y|$, respectively.\\

\noindent
\emph{Proof.}
For the first inequality, we define 
\begin{equation}
	\rho(x) = \arcoth[x] - \frac{1}{y}\arcoth\bx\frac{x}{y}\by
		= \frac{1}{2}\ln\bx\biggl(\frac{x+1}{x-1}\biggr)\biggl(\frac{x-y}{x+y}\biggr)^{\frac{1}{y}}\by
\end{equation}
and note that 
\begin{equation}
\rho'(x) = \frac{y^2 -1}{(x^2-1)(x^2 - y^2)}.
\end{equation}
Hence, $\rho(x)$ is non-increasing in $x\geq 1$ for any $|y|\leq 1$. 
Since $\lim_{x\rightarrow\infty} \rho(x) =0$, it follows that $\rho(x)\geq 0$. 
The second inequality can be proven analogously.

\end{document}